\documentclass[runningheads]{llncs}
\usepackage[T1]{fontenc}
\usepackage{graphicx,verbatim}

\usepackage{verbatim}
\usepackage{graphicx}
\usepackage{bibentry}
\usepackage{amsmath}
\usepackage{multirow}
\usepackage{booktabs}
\usepackage{tabularx}
\usepackage{makecell}
\usepackage{bbding}
\usepackage{pifont}
\usepackage{wasysym}
\usepackage{amssymb}
\usepackage{svg}
\usepackage[table]{xcolor}
\usepackage{rotating}
\usepackage[export]{adjustbox}
\usepackage{anyfontsize}

\newcommand{\figref}[1]{Fig.~\ref{#1}}

\newcommand{\tabref}[1]{Tab.~\ref{#1}}
\newcommand{\eqnref}[1]{Eq.~(\ref{#1})}

\newcommand{\Figref}[1]{Figure~\ref{#1}}

\newcommand{\Tabref}[1]{Table~\ref{#1}}

\newcommand{\ie}{\textit{i.e.}}

\usepackage{hyperref}
\definecolor{watermelon_green}{rgb}{0, 0.42, 0.30}

\begin{document}
\title{Zero-shot CT Super-Resolution using Diffusion-based 2D Projection Priors and \\ Signed 3D Gaussians}

\titlerunning{Zero-shot CT SR using Diffusion-based SR Proj. and NAB-GS}
\begin{comment}  %% Removed for anonymized MICCAI submission
\author{First Author\inst{1}\orcidID{0000-1111-2222-3333} \and
Second Author\inst{2,3}\orcidID{1111-2222-3333-4444} \and
Third Author\inst{3}\orcidID{2222--3333-4444-5555}}
%
\authorrunning{F. Author et al.}
\institute{Princeton University, Princeton NJ 08544, USA \and
Springer Heidelberg, Tiergartenstr. 17, 69121 Heidelberg, Germany
\email{lncs@springer.com}\\
\url{http://www.springer.com/gp/computer-science/lncs} \and
ABC Institute, Rupert-Karls-University Heidelberg, Heidelberg, Germany\\
\email{\{abc,lncs\}@uni-heidelberg.de}}

\end{comment}

\author{Jeonghyun Noh$^*$, Hyun-Jic Oh$^*$, Won-Ki Jeong$^\dagger$}  %% Added for anonymized MICCAI submission
\authorrunning{Noh et al.}
\institute{Department of Computer Science and Engineering, Korea University, Seoul, Korea \\
    \email{\{wjdgus0967, hyunjic0127, wkjeong\}@korea.ac.kr}}

% \author{Anonymized Authors}  %% Added for anonymized MICCAI submission
% \authorrunning{Anonymized Author et al.}
% \institute{Anonymized Affiliations \\
%     \email{email@anonymized.com}}
  
\maketitle              % typeset the header of the contribution
\def\thefootnote{$*$}\footnotetext{Co-first authors.}
\def\thefootnote{$\dagger$}\footnotetext{Corresponding author.}

\begin{abstract}
Computed tomography (CT) is important in clinical diagnosis, but acquiring high-resolution (HR) CT is constrained by radiation exposure risks.
While deep learning-based super-resolution (SR) methods have shown promise for reconstructing HR CT from low-resolution (LR) inputs, supervised approaches require paired datasets that are often unavailable.
Zero-shot methods address this limitation by operating on single LR inputs; however, they frequently fail to recover fine structural details due to limited LR information within individual volumes.
To overcome these limitations, we propose a novel zero-shot 3D CT SR framework that integrates diffusion-based upsampled 2D projection priors into the 3D reconstruction process. % diffusion-based upsampled 2D projection prior
Specifically, our framework consists of two stages: (1) LR CT projection SR, training a diffusion model on abundant X-ray data to upsample LR projections, thereby enhancing the scarce information inherent in the LR inputs.
(2) 3D CT volume reconstruction, using 3D Gaussian splatting with our novel Negative Alpha Blending (NAB-GS), which models positive and negative Gaussian densities to learn signed residuals between diffusion-generated HR and upsampled LR projections.
Our framework demonstrates superior quantitative and qualitative performance on two public datasets, and expert evaluations present the framework's clinical potential at 4$\times$.
\href{https://github.com/hvcl/3DCT_SR_DDNM_NAB-GS}{Github-code}.

\keywords{Zero-shot CT Super-resolution \and Diffusion Models \and 3D Gaussian Splatting.}
% Authors must provide keywords and are not allowed to remove this Keyword section.

\end{abstract}

\section{Introduction}
\label{sec:intro}
High-quality computed tomography (CT) is an essential modality for accurate clinical diagnosis and treatment planning.
However, acquiring high-resolution (HR) CT scans necessitates elevated radiation doses, risking potential DNA damage and radiation-induced malignancies~\cite{martin2006health}.
While reducing radiation dose mitigates these concerns, it inevitably limits the data available for image reconstruction, thereby degrading spatial resolution~\cite{mccollough2015degradation}.
This trade-off often obscures anatomical structures, potentially compromising diagnostic precision.

To address this, super-resolution (SR) techniques have emerged as a vital tool for recovering detailed structures from low-resolution (LR) inputs~\cite{frazer2024super}.
Deep learning has revolutionized a wide range of computer vision tasks~\cite{gharib2025exploring,noh2025dual,oh2023diffmix}.
In particular, SR has benefited from advances in convolutional neural networks~\cite{dong2015image,zhang2018residual}, transformers~\cite{chen2023activating,liang2021swinir}, and diffusion models~\cite{kawar2022denoising,liu2023accelerating,saharia2022image}, these are primarily tailored to 2D images and often fail to exploit volumetric consistency in medical imaging.
In the medical domain, supervised 3D SR approaches, ranging from spatially aware interpolation networks~\cite{peng2020saint} to implicit neural representations (INR)~\cite{fang2024cycleinr,wu2022arbitrary}, have been proposed.
Despite such effectiveness, applicability is limited by scarce paired HR-LR training volumes.
Consequently, zero-shot learning has gained traction as a data-efficient alternative~\cite{shocher2018zero,soh2020meta,wang2022uassr}, with recent innovations like CuNeRF~\cite{chen2023cunerf} adapting neural radiance fields for continuous upsampling.
Yet, relying solely on LR information, current zero-shot methods often struggle to recover structural details, yielding over-smoothed reconstructions.

To overcome these limitations, we propose a novel zero-shot 3D CT SR framework that reformulates volumetric SR as a reconstruction task using diffusion-based upsampled 2D projection priors.
The proposed framework consists of two key stages: (1) LR CT projection SR, and (2) 3D reconstruction.
First, given that acquiring 2D X-ray data is more feasible than collecting paired 3D HR-LR volumes, we adopt a diffusion model trained on large-scale 2D X-ray datasets.
We then employ this diffusion prior within a Denoising Diffusion Null-space Model (DDNM)~\cite{wang2022zero} to generate 2D HR CT projections from LR counterparts.
This design choice intends LR projections to enforce data consistency, simultaneously enhancing the projection information through the diffusion model prior.
Second, we introduce a Negative Alpha Blending Gaussian Splatting (NAB-GS) to reconstruct the HR 3D CT volume.
NAB-GS learns the residual field between the diffusion-generated HR projections and the projections of the upsampled LR volume.
To fully exploit this residual field, which inherently contains both positive and negative values, NAB-GS relaxes the non-negativity constraint of standard 3DGS~\cite{kerbl20233d}, enabling precise encoding of signed residuals and substantially improving the recovery of structural details.
Consequently, our framework achieves superior results on two public datasets, with expert evaluations suggesting its clinical potential at 4$\times$.
Our main contributions are summarized as follows:

\begin{itemize}
    \item We propose a novel zero-shot 3D CT SR framework that reformulates volumetric SR as a 3D reconstruction task driven by diffusion-based upsampled 2D projection priors.
    Leveraging a diffusion model trained on large-scale X-ray datasets, we mitigate the need for paired HR-LR volumes and enhance the limited information of LR inputs.
    \item We introduce NAB-GS, a Gaussian splatting that relaxes the non-negativity constraint of standard 3DGS to learn signed residual fields between diffusion-generated HR projections and upsampled LR projections, enabling precise residual encoding and improved structural details.
    \item We validate our framework on the UHRCT~\cite{chu2023topology} and MELA\footnote[1]{https://mela.grand-challenge.org/} public datasets, achieving superior quantitative and qualitative performance compared to state-of-the-art (SOTA) zero-shot methods, alongside expert evaluations supporting its CT volume quality and clinical potential.
\end{itemize}

\section{Zero-Shot 3D CT SR Framework}
\begin{figure}[t]
\centering
\includegraphics[width=0.99\textwidth]{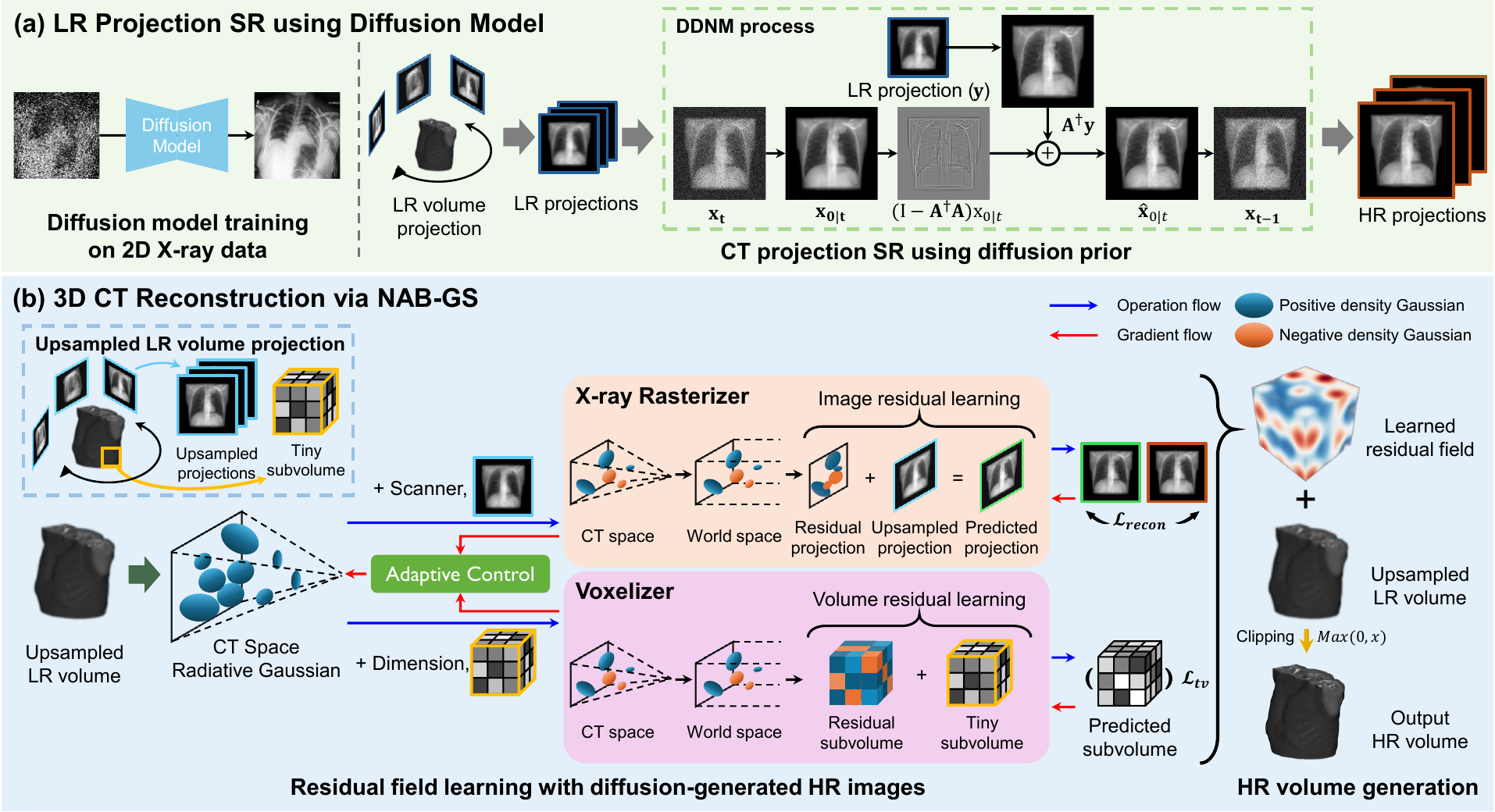}
\centering
\caption{
% \HJ{Bigger the text sizes}
% Overview of the proposed zero-shot 3D CT SR framework. (a) 2D X-ray Projection SR: LR 2D X-ray projection images are derived from the LR volume and upsampled using DDNM with PAS. (b) 3D CT Reconstruction: The 3D volume is reconstructed using upsampled projection images using NAB-GS.
% Overview of the proposed zero-shot 3D CT SR framework. (a) 2D X-ray Projection SR: LR 2D X-ray projections are derived from the input LR volume and upsampled using DDNM with PAS. (b) 3D CT Reconstruction: The HR volume is reconstructed from the upsampled projections via NAB-GS, which learns residual fields to improve anatomical details.
Overview of our framework. (a) LR projection SR using diffusion model: A pre-trained diffusion model with 2D X-ray data is employed within the DDNM to generate HR 2D CT projection images from LR counterparts. 
(b) 3D CT reconstruction via NAB-GS:
% It learns a signed residual field, using both positive and negative density Gaussians, with diffusion-generated HR images. For HR volume generation, the learned residual field is added onto the upsampled LR volume.
% \HJ{
Using both positive and negative density Gaussians, we model a signed residual field between diffusion-generated HR projections and LR counterparts.
For HR volume generation, the learned residual field is added onto the upsampled LR volume.
% }
}
\label{fig:overall_architecture}
\end{figure}

% \label{sec:method}
\label{sec:method_framework}
\subsection{LR Projection SR using Diffusion Model}
\label{sec:2D_SR}

% As depicted in~\ref{fig:overall_architecture} (a), we upsample 2D X-ray projection using the pre-trained diffusion model which has prior knowledge on 2D X-ray domain using DDNM.
% As illustrated in Fig.~\ref{fig:overall_architecture} (a), we perform DDNM-based~\cite{wang2022zero} 2D X-ray projection SR using a pre-trained diffusion model as a domain prior.

% The first step is %
%As introduced in~\Secref{sec:method_framework}, we perform 
The first stage focuses on generating high-fidelity 2D CT projections to serve as reliable guidance for 3D reconstruction, as shown in~\figref{fig:overall_architecture}(a).
% We utilize a diffusion model pre-trained on a large-scale dataset of 2D X-ray images, leveraging it as a robust domain prior to generate high-frequency details from LR inputs.
%
% We train a diffusion model on large-scale 2D X-ray datasets, learning a generative prior that captures realistic projection anatomy and contrast.
% We train a diffusion model on large-scale 2D X-ray datasets, learning a generative projection prior.
% \JH{
We train a diffusion model on large-scale 2D X-ray datasets specifically to model a generative projection prior.
% }
%
% Subsequently, this diffusion prior is adapted to our CT projection SR task using DDNM~\cite{wang2022zero}, enabling HR CT projection generation from LR CT projections with data-consistency constraints.
Subsequently, this prior is adapted to our CT projection SR task using DDNM~\cite{wang2022zero}, enhancing the limited information in LR CT projections.
% To apply this prior to our SR task with no additional training, we adopt the DDNM~\cite{wang2022zero} framework.
% for projection-wise SR, as illustrated in~\figref{fig:overall_architecture}(a).
% Specifically, as illustrated in~\figref{fig:overall_architecture}(a), we employ a DDNM-based approach~\cite{wang2022zero}, leveraging the diffusion prior for projection-wise SR.
% In this stage, we perform DDNM-based~\cite{wang2022zero} 2D X-ray projection super-resolution using a pre-trained diffusion model as a domain prior (see~\figref{fig:overall_architecture}(a)).

\subsubsection{SR using DDNM.} 
% \HJ{
% % (1) inverse A, downsampling operator information 
% (2) X-ray trained diffusion -> CT projection SR can improve LR information (Table 2: Ours vs Bilinear).
% }
% Our upsampling process relies on DDNM.
% Our 2D X-ray projection upsampling process relies on DDNM.
% Let $\mathbf{y} = \mathbf{A}\mathbf{x}$ denote an LR CT projection, where $\mathbf{x}$ is the unknown HR projection and $\mathbf{A}$ is the degradation operator (\eg, downsampling).
Let $\mathbf{y} = \mathbf{A}\mathbf{x}$ denote an LR CT projection, where $\mathbf{x}$ is the unknown HR projection and $\mathbf{A}$ is the downsampling operator. % (\eg, downsampling).
Our goal is to estimate an HR projection $\hat{\mathbf{x}}$ that both matches the measured LR projection and lies on the manifold modeled by the diffusion prior.

Following the null-space based diffusion formulation in~\cite{wang2022zero}, we combine the learned prior with a data-consistency constraint defined by $\mathbf{y}$.
% At each reverse diffusion step, the current estimate is decomposed into range and null-space components:
At each reverse diffusion step, the estimate is decomposed into range and null-space components:
\begin{equation}
\hat{\mathbf{x}}_{0|t} = \mathbf{A}^{\dagger}\mathbf{y} + (\mathbf{I} - \mathbf{A}^{\dagger}\mathbf{A}) \mathbf{x}_{0|t},
\end{equation}
where $\mathbf{A}^{\dagger}$ denotes the pseudo-inverse of $\mathbf{A}$ and $\mathbf{x}_{0|t}$ is the denoised estimate predicted by the diffusion model at step $t$.
% The first term enforces data consistency in the range space of $\mathbf{A}$, while the second term allows the diffusion prior to refine details in the null space, injecting high-frequency structures that are not directly constrained by $\mathbf{y}$.
The first term enforces data consistency in the range space of $\mathbf{A}$, while the second term allows the diffusion prior to refine details in the null space, injecting high-frequency structures. % not in $\mathbf{y}$.

To account for measurement noise or slight inconsistencies in LR projections, we further adopt the DDNM+~\cite{wang2022zero}:
% To account for measurement noise or slight inconsistencies inherent in the LR inputs,
% DDNM+~\cite{wang2022zero}, which introduces a scaled correction term:
\begin{equation}
\hat{\mathbf{x}}_{0|t} = \mathbf{x}_{0|t} - \Sigma_t \mathbf{A}^{\dagger}(\mathbf{A}\mathbf{x}_{0|t} - \mathbf{y}),
\end{equation}
where $\Sigma_t$ controls the strength of the correction at each diffusion step, balancing data fidelity and perceptual quality.
This null-space diffusion process allows us to generate HR CT projections that inherit realistic projection appearance from the trained diffusion prior while leveraging LR projection information to enforce data consistency.
In this way, an X-ray trained diffusion prior coupled with DDNM offers a practical design to enhance the limited information in LR CT inputs under a zero-shot SR setting.

\subsection{3D CT Reconstruction via NAB-GS}
\label{sec:NAB-GS}

% \wk{explain whole 3D process more.. with residual learning}
% Using upsampled 2D X-ray images, we reconstruct the 3D CT volume through 3DGS, as described in \figref{fig:overall_architecture}(b).
The second stage performs 3D CT reconstruction via the proposed NAB-GS, as shown in~\figref{fig:overall_architecture}(b).
%
% \HJ{cubic interpolation / cubic interpolation}
% The process begins by initializing a radiative Gaussian field from the upsampled LR volume using cubic interpolation.
%
% As shown in~\figref{fig:overall_architecture}(b), the process begins by initializing the radiative Gaussian field from the upsampled LR volume using cubic interpolation.
We obtain an initial upsampled LR volume by cubic upsampling of the LR volume and initialize a radiative Gaussian field from this upsampled LR volume.
%
% Rather than directly fitting the full HR volume, we formulate reconstruction as learning a residual field between the diffusion-generated HR projections and the projections of the upsampled LR volume.
% We reconstruct the HR volume by learning a residual field between the diffusion-generated HR projections and the upsampled LR baseline, enhancing anatomical details.
To reconstruct the HR volume, we learn a residual field between the diffusion-generated HR projections and the upsampled LR baseline.
%
% In parallel, a projection and a corresponding tiny subvolume are generated from the same upsampled LR volume to facilitate residual learning.
% %
% These auxiliary representations guide the optimization of the Gaussian parameters through an X-ray rasterizer and voxelizer, respectively.
% %
% For effective training, we employ reconstruction and total variation (TV) losses, where the reconstruction loss is computed between the predicted and the diffusion-based upsampled image.
% We then optimize it using an X-ray rasterizer and voxelizer, leveraging auxiliary projections and tiny subvolumes, generated from an upsampled LR volume, to guide residual learning under reconstruction and total variation (TV) losses.
% We optimize the Gaussian parameters using an X-ray rasterizer and voxelizer under reconstruction and total variation (TV) losses, so that the learned residuals refine the baseline volume while preserving consistency across views.
The Gaussians are optimized via rasterizer and voxelizer (following R$^2$-GS~\cite{ruyi2024r2_gaussian}) under reconstruction and total variation (TV) losses to refine structural details.

% \subsubsection{Residual learning.}
\subsubsection{Negative alpha blending.}
Conventional 3DGS~\cite{kerbl20233d} enforces non-negativity of density $\rho$ by softplus activation ($\rho = \ln(1 + e^z)$) and alpha blending, where $z$ is the raw density.
While physically plausible, this restricts the representation to strictly positive densities.
However, our residual formulation targets the discrepancy between the diffusion-generated HR projection and the upsampled LR baseline.
Since the upsampled LR can locally over- or underestimate true intensities, the residual inherently contains both positive and negative values.

To faithfully encode signed residuals, first, we replace the softplus with Parametric ReLU (PReLU), allowing for negative densities:
\begin{align}
    \phi(z) = 
    \begin{cases}
    z, & \text{if } z \geq 0 \\
    \gamma z, & \text{otherwise}
    \end{cases} \; ,
    \;\;\;
    \phi^{-1}(\rho) = 
    \begin{cases}
    \rho, & \text{if } \rho \geq 0 \\
    \frac{\rho}{\gamma}, & \text{otherwise}
    \end{cases} \;,
\end{align}
where $\phi(\cdot)$ and $\phi^{-1}(\cdot)$ denote the PReLU and its inverse, and $\gamma$ is a learnable negative-slope parameter that controls the gradient magnitude.
This regulation prevents divergence, stabilizing the optimization process across the residual field.

While $z$ can now take negative values, standard alpha blending still constrains the accumulation to be non-negative for physical plausibility.
Thus, we modify the rendering formulation to permit negative contributions within the blending process.
Following R$^2$-GS, we model the linear integral of X-ray projection without the standard transmittance term:
\begin{equation}
    \label{eq:alpha_blending}
    C = \sum_{i=1}^{N} \alpha_i, \quad \text{subject to} \; (\alpha_i \geq \epsilon),
\end{equation}
where $\alpha_i$ and $\epsilon$ denote the $i$-th Gaussian's contribution and a heuristic threshold, respectively.
To actively accommodate negative values, we specifically remove the strict non-negativity filter ($\alpha_i \geq \epsilon$), conventionally imposed to ensure physical plausibility, from \eqnref{eq:alpha_blending}.
%
% Despite this change, our linear accumulation remains stable, inherently resolving occlusion issues and ensuring constant gradients ($\frac{\partial C}{\partial \alpha_i} = 1$).
Despite this change, since our formulation is a linear accumulation, the gradient remains constant ($\frac{\partial C}{\partial \alpha_i} = 1$) regardless of whether $\alpha_i$ is positive or negative, ensuring stable optimization.
Accordingly, we adapt our pruning process, discarding Gaussians only if $|z| < 10^{-5}$.
Finally, the output HR volume is reconstructed by adding the learned residual field to the upsampled LR volume, followed by clipping ($\mathtt{max}(0, \cdot)$) to ensure physical plausibility.
This mechanism, NAB-GS, can selectively amplify or suppress local intensities of the upsampled LR volume, successfully recovering fine structure details.

\noindent\textbf{Loss function.}
% To optimize 3D CT reconstruction, we define a total loss function, defined as:
% \HJ{Define/define redundancy. define a loss function vs compose our loss function?}
To optimize 3D CT reconstruction, the total loss function is:
% To optimize 3D CT reconstruction, we define a total loss function that combines various loss terms, defined as:
%
% % \begin{equation}
% \begin{align}
%     \label{equ:loss}
%      &\mathcal{L}_{recon} = \mathcal{L}_1(y,x) + \mathcal{L}_{res}(\hat{y}, \hat{x}) + \lambda_1 \mathcal{L}_{dssim}(y,x), \\
%      &\mathcal{L}_{total} = \mathcal{L}_{recon} +  \lambda_2 \mathcal{L}_{tv},
% \end{align}
% % \end{equation}
\begin{equation}
     \mathcal{L}_{total} = \mathcal{L}_{recon} + \lambda_1 \mathcal{L}_{tv}, \; \; \mathcal{L}_{recon} = \mathcal{L}_1(y,x) + \mathcal{L}_{res}(\hat{y}, \hat{x}) + \lambda_2 \mathcal{L}_{ssim}(y,x),
\end{equation}
%
% \HJ{GT: is it diffusion output? need to give a clearer explanation}
where $y$ and $x$ denote the diffusion-generated HR projection and our predicted projection, while $\hat{y}$ and $\hat{x}$ represent residual projection (\ie, $y \ -$ upsampled projection) and residual projection.
The terms $\mathcal{L}_1(\cdot,\cdot)$, $\mathcal{L}_{res}(\cdot,\cdot)$, $\mathcal{L}_{ssim}(\cdot,\cdot)$, and $\mathcal{L}_{tv}(\cdot)$ correspond to the L1, L1-based residual, structural similarity index measure (SSIM), and TV loss function, respectively.
The hyperparameters are empirically set to $\lambda_1 = 0.05$ and $\lambda_2 = 0.5$.
Following R$^2$-GS, we compute $\mathcal{L}_{tv}(\cdot)$ efficiently using randomly cropped 32$^3$ subvolumes during training.

% where $y$, $x$, $\hat{y}$, and $\hat{x}$ represent the diffusion output, the prediction, the residual GT (\ie, GT minus the upsampled LR), and the residual prediction (\ie, before addition with the upsampled LR), respectively.
%
% $\mathcal{L}_1(\cdot,\cdot)$, $\mathcal{L}_{res}(\cdot,\cdot)$, $\mathcal{L}_{ssim}(\cdot,\cdot)$, and $\mathcal{L}_{tv}(\cdot,\cdot)$ denote L1 loss, residual loss based on L1, structural similarity index measure (SSIM) loss, and TV loss, respectively.
%
% The hyperparameters $\lambda_1$ and $\lambda_2$ control the loss weights and are set to 0.05 and 0.5, respectively.
%
% \JH{
% Note that, following R$^2$-GS, we use randomly cropped 32$^3$ volumes during training to compute $\mathcal{L}_{tv}$ efficiently.
%\HJ{L -> mathcal L}
% }
% \wk{Doesn't this statement apply only to $L_{tv}$ loss ?}
%
%The total loss $\mathcal{L}_{total}$ enables NAB-GS to be trained in an efficient and stable.

\section{Experiments}

\subsection{Setup}
\label{sec:setup}
\noindent\textbf{Dataset.}
\noindent\textbf{[2D]}
We train our unconditional diffusion model on two large-scale X-ray datasets resized to 512$\times$512.
ChestX-ray14\footnote[2]{https://nihcc.app.box.com/v/ChestXray-NIHCC} (112,120 frontal images) and CheXpert\footnote[3]{https://stanfordmlgroup.github.io/competitions/chexpert/} (80,845 frontal and lateral images).
\noindent\textbf{[3D]}
Evaluation is performed on UHRCT~\cite{chu2023topology} (20/10 train/test split) and MELA (60/20/20 train/val/test split).
Volumes are resized to 512$^3$, clipped to [-512, 3071], and normalized to [0, 1].
LR volumes are generated via Gaussian smoothing followed by sinc interpolation.
Note that our zero-shot learning uses only test sets.

\noindent\textbf{Implementation details.}
\textbf{[Diffusion]} is conducted in PyTorch 1.11.0. % \HJ{Todo: DDPM setup}
We trained an unconditional DDPM~\cite{ho2020denoising} using four NVIDIA A6000 GPUs.
For inference, we used DDIM~\cite{song2020denoising} sampling with 50 steps, employing DDNM+~\cite{wang2022zero} for both 4$\times$ and 8$\times$ SR with a noise level of $0.0015$.
We upscaled 100 X-ray projections, uniformly spaced between 0$^{\circ}$ to 180$^{\circ}$, from each LR volume using LR projection by TIGRE~\cite{biguri2016tigre}.
\textbf{[NAB-GS]} is implemented in PyTorch and trained for 5k iterations on a single RTX 3090 GPU.
We adopt the default training parameter of R$^2$-GS~\cite{ruyi2024r2_gaussian}, except at 8$\times$, where the learning rates for scaling (from $5 \times 10^{-4}$ to $5 \times 10^{-5}$) and rotation (from $1 \times 10^{-4}$ to $1 \times 10^{-5}$).
Our framework takes around 15 minutes per volume (Diffusion 10 min, NAB-GS 5 min).
Performance is evaluated using PSNR and SSIM~\cite{wang2004image}.

\begin{table}[t]
\caption{
Quantitative comparison of 3D CT SR with previous methods.
}
\label{tab:3DCT_results}
\centering
% \resizebox{0.95\textwidth}{!}
{
\fontsize{8.1pt}{9.5pt}\selectfont
\begin{tabular}{l|cc|cc|cc|cc}
\Xhline{2\arrayrulewidth}
\multirow{2}{*}{\textbf{Methods}}
& \multicolumn{2}{c|}{\textbf{UHRCT 4$\times$}} 
& \multicolumn{2}{c|}{\textbf{UHRCT 8$\times$}} 
& \multicolumn{2}{c|}{\textbf{MELA 4$\times$}} 
& \multicolumn{2}{c}{\textbf{MELA 8$\times$}} 

\\ %\cline{5-10}

% & \textbf{PSNR}$\uparrow$ & \textbf{SSIM}$\uparrow$ 
% & \textbf{PSNR}$\uparrow$ & \textbf{SSIM}$\uparrow$ 
% & \textbf{PSNR}$\uparrow$ & \textbf{SSIM}$\uparrow$ 
% & \textbf{PSNR}$\uparrow$ & \textbf{SSIM}$\uparrow$ 
& PSNR$\uparrow$ & SSIM$\uparrow$
& PSNR$\uparrow$ & SSIM$\uparrow$
& PSNR$\uparrow$ & SSIM$\uparrow$
& PSNR$\uparrow$ & SSIM$\uparrow$
\\ \hline \hline

Trilinear
& 24.56 & \underline{0.8877}
& 21.14 & \underline{0.8125}
& 33.64 & \underline{0.9472}
& 30.27 & \underline{0.9063} \\

Cubic
& 24.61 & 0.8783
& \underline{21.46} & 0.8013
& 33.55 & 0.9431
& \underline{30.49} & 0.9005 \\

% \hline

% ArSSR~\cite{wu2022arbitrary}
% & 32.92 & \textbf{0.9589}
% & \underline{30.38} & \textbf{0.9294}
% & 24.81 & \underline{0.8900}
% & \underline{21.72} & \textbf{0.8458} \\

\hline

NeRF~\cite{mildenhall2021nerf}
& 20.86 & 0.6745 
& 18.92 & 0.5797 
& 29.76 & 0.8088 
& 28.73 & 0.7873 \\

CuNeRF~\cite{chen2023cunerf}
& \underline{25.25} & 0.8459
& 21.04 & 0.7572
& \underline{33.76} & 0.9096
& 30.11 & 0.8535 \\
% \hline

% Bilinear $+$ NAB-GS
% & 24.57 & 0.8244
% & 21.34 & 0.8015
% & 33.50 & 0.8963
% & 30.42 & 0.9009 \\

% RDN $+$ NAB-GS
% & 23.78 & 0.6410
% & 21.41 & 0.6946
% & 31.92 & 0.7903
% & 30.27 & 0.8472 \\

% Omni $+$ NAB-GS
% & 22.74 & 0.5621
% & 20.56 & 0.5992
% & 31.84 & 0.7732
% & 29.32 & 0.7771 \\

% \textbf{PAS $+$ NAB-GS}
\textbf{Ours}
& \textbf{25.42} & \textbf{0.8957}
& \textbf{21.96} & \textbf{0.8172}
& \textbf{34.17} & \textbf{0.9525}
& \textbf{30.81} & \textbf{0.9115} \\

\hline
% (Supervised) ArSSR~\cite{wu2022arbitrary}
Supervised (ArSSR~\cite{wu2022arbitrary})
% Supervised~\cite{wu2022arbitrary}
& 22.78 & 0.8936
& 21.10 & 0.8336
& 33.00 & 0.9658
& 30.70 & 0.9343 \\

\Xhline{2\arrayrulewidth}
% \multicolumn{13}{r}{\textbf{Bold}: The best, \underline{Underline}: The second-best}
\end{tabular}
}
\end{table}

\begin{figure}[h!]
\centering
\includegraphics[width=0.95\textwidth]{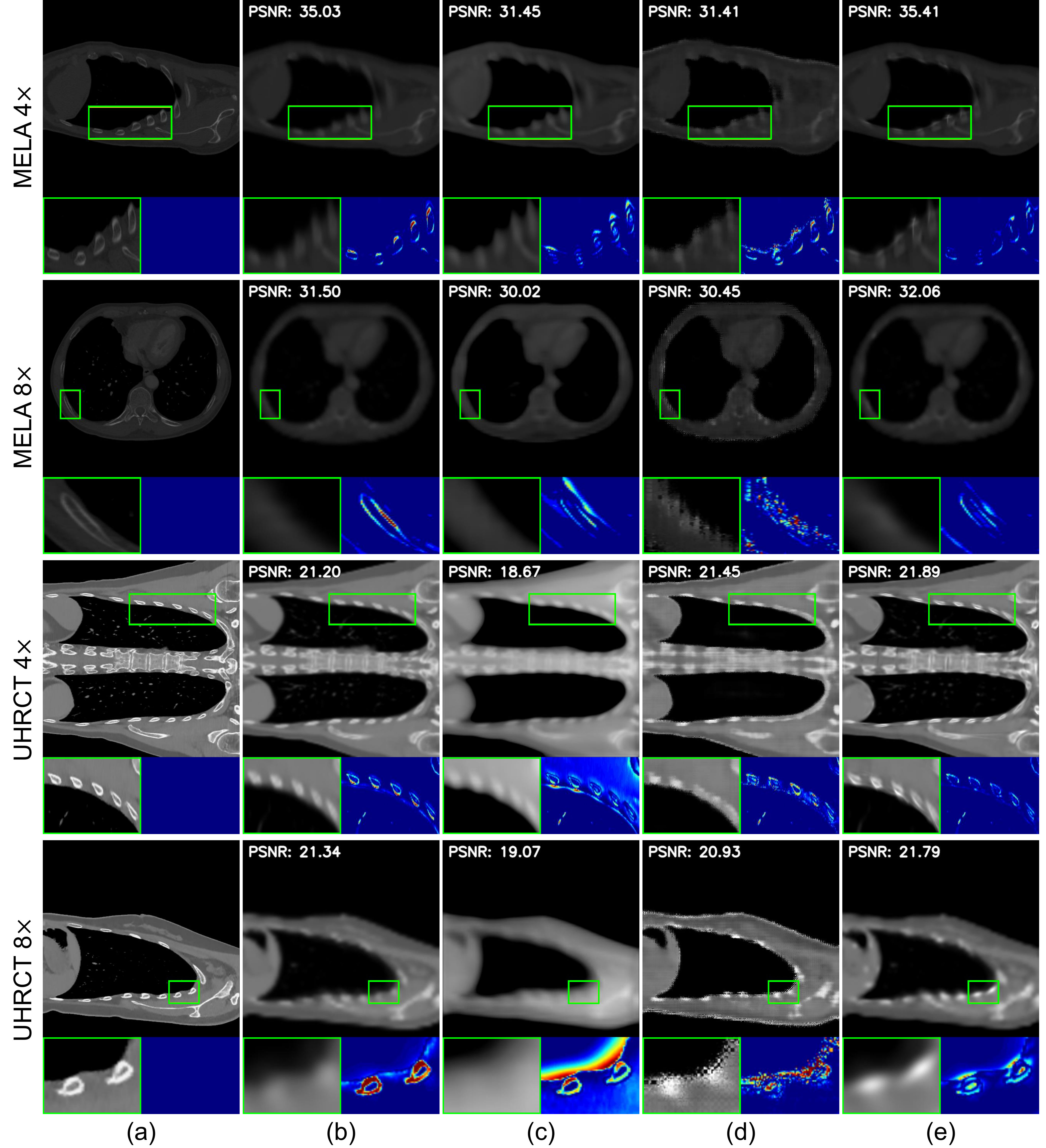}
\centering
\caption{
    Visual comparisons of 3D CT reconstruction results.
    (a) Ground truth (GT), (b) Cubic interpolation, (c) ArSSR~\cite{wu2022arbitrary}, (d) CuNeRF~\cite{chen2023cunerf}, and (e) Ours.
    The green box and zoom-in highlight regions where our method excels at reconstruction.
    The error map is computed by the L2 norm between the prediction and the ground truth.
}
\label{fig:visual_results}
\end{figure}

\subsection{Results and Evaluation}
\noindent\textbf{Quantitative results.}
\Tabref{tab:3DCT_results} presents the comparison against conventional upsampling (Trilinear, Cubic), zero-shot (NeRF~\cite{mildenhall2021nerf}, CuNeRF~\cite{chen2023cunerf}), and supervised (ArSSR~\cite{wu2022arbitrary}) methods, where CuNeRF is the only publicly available zero-shot baseline for 3D CT SR.
ArSSR was trained on the train set of each dataset; NeRF, implemented as in CuNeRF, used standard ray sampling for SR, unlike CuNeRF's cube sampling.
Overall, our method achieves superior PSNR and SSIM across both the UHRCT and MELA datasets.
Notably, when compared to CuNeRF, which relies solely on internal volume information, our approach yields substantially higher SSIM (+0.04-0.06).
Compared to ArSSR, our zero-shot framework demonstrates competitive performance.
Furthermore, our framework is computationally efficient, completing in 15 minutes, whereas CuNeRF takes about 1 hour and ArSSR takes about 25 minutes per volume. %, respectively.

\noindent\textbf{Qualitative results.}
\Figref{fig:visual_results} visualizes the slice of reconstructed HR volumes and their corresponding L2 error maps. %
Cubic (b) suffers from severe over-smoothing, and CuNeRF (d) introduces noticeable high-frequency artifacts, particularly at 8$\times$.
In contrast, our method (e) yields more faithful structural details.
Notably, at 4$\times$ upsampling (rows 1 and 3), ours more precisely restores fine structures such as bone boundaries.
These quantitative and qualitative results clearly demonstrate the effectiveness of the proposed method.

\noindent\textbf{Experts' evaluation.}
A user study with two domain experts was conducted to assess the clinical potential of the proposed method, comparing MELA (4$\times$, 8$\times$) volumes produced by three methods (Cubic, CuNeRF, and Ours).
The experts commented that (1) our method at 4$\times$ shows potential for real-world clinical use, while further improvements are required at 8$\times$ and (2) ours yields sharper results than Cubic, but the current visual gain does not substantially contribute to overall utility.
They further noted that (3) it clearly outperforms CuNeRF by more accurately preserving fine structural details and (4) enhancing inter-slice consistency is an important direction for further improving clinical utility.

\begin{table}[t]
    \centering
    \begin{minipage}{.47\linewidth}
        \caption{Ablation on 2D projection SR methods on 3D CT SR performance.}
        \label{tab:ablation_2d}
        \centering
        \resizebox{\textwidth}{!}
        {
        % \fontsize{8.5pt}{9.5pt}\selectfont
        \begin{tabular}{l|cc|cc}
        \Xhline{2\arrayrulewidth}
        \multirow{2}{*}{\textbf{Methods}}
        & \multicolumn{2}{c|}{\textbf{UHRCT 4$\times$}} 
        & \multicolumn{2}{c}{\textbf{UHRCT 8$\times$}} 
        \\
        % & \textbf{PSNR}$\uparrow$ & \textbf{SSIM}$\uparrow$ 
        % & \textbf{PSNR}$\uparrow$ & \textbf{SSIM}$\uparrow$ 
        & PSNR$\uparrow$ & SSIM$\uparrow$
        & PSNR$\uparrow$ & SSIM$\uparrow$
        \\ \hline \hline
        Bilinear
        & 24.57 & 0.8244
        & 21.34 & 0.8015 \\

        RDN~\cite{zhang2018residual}
        & 23.78 & 0.6410
        & 21.41 & 0.6946 \\
        
        Omni~\cite{wang2023omni}
        & 22.74 & 0.5621
        & 20.56 & 0.5992 \\
        
        \hline
        
        BIRD~\cite{chihaoui2024blind}
        & 20.18 & 0.8111
        & 19.62 & 0.7845 \\ 

        GDP~\cite{fei2023generative}
        & 22.84 & 0.8485
        & 21.00 & 0.8075 \\

        \hline

        \textbf{Ours}
        & \textbf{25.42} & \textbf{0.8957}
        & \textbf{21.96} & \textbf{0.8172} \\
        \Xhline{2\arrayrulewidth}
        \end{tabular}}
    \end{minipage}
    \begin{minipage}{.51\linewidth}
        \caption{Ablation on 3D reconstruction methods and activation functions.}\label{tab:ablation_3d}
        \centering
        \resizebox{\textwidth}{!}
        {
        % \fontsize{8.5pt}{9.5pt}\selectfont
        \begin{tabular}{l|cc|cc}
        \Xhline{2\arrayrulewidth}
        \multirow{2}{*}{\textbf{Methods}}
        & \multicolumn{2}{c|}{\textbf{UHRCT 4$\times$}} 
        & \multicolumn{2}{c}{\textbf{UHRCT 8$\times$}} 
        \\
        % & \textbf{PSNR}$\uparrow$ & \textbf{SSIM}$\uparrow$ 
        % & \textbf{PSNR}$\uparrow$ & \textbf{SSIM}$\uparrow$ 
        & PSNR$\uparrow$ & SSIM$\uparrow$
        & PSNR$\uparrow$ & SSIM$\uparrow$
        \\ \hline \hline
        FDK~\cite{feldkamp1984practical}
        & 21.88 & 0.5375 
        & 19.39 & 0.4337 \\
        
        NAF~\cite{zha2022naf}
        & 24.23 & 0.8559 
        & 21.66 & 0.7982 \\
        
        SAX-NeRF~\cite{cai2024structure}
        & 24.81 & 0.8719 
        & 21.40 & 0.7826 \\
        
        R$^2$-GS~\cite{ruyi2024r2_gaussian}
        & 24.89 & 0.8839 
        & 21.25 & 0.8056 \\ \hline
        
        Softplus
        & 24.71 & 0.8790 
        & 21.61 & 0.8022 \\

        ReLU
        & 24.62 & 0.8784
        & 21.57 & 0.8021 \\
        
        Sine
        & 25.19 & 0.8475 
        & 21.26 & 0.7208 \\

        \hline

        \textbf{Ours}
        & \textbf{25.42} & \textbf{0.8957}
        & \textbf{21.96} & \textbf{0.8172} \\
        
        \Xhline{2\arrayrulewidth}
        \end{tabular}}
    \end{minipage}
\end{table}

\noindent\textbf{Ablation on UHRCT.}
\noindent\textbf{(1) 2D projection SR.}
We evaluate the impact of 2D SR methods on 3D SR performance in~\tabref{tab:ablation_2d}.
We compare our method against non-diffusion baselines, including bilinear interpolation, RDN~\cite{zhang2018residual}, and Omni~\cite{wang2023omni}, as well as diffusion-based optimization methods, including GDP~\cite{fei2023generative} and BIRD~\cite{chihaoui2024blind}.
RDN and Omni were trained on projections generated from the UHRCT training set via TIGRE.
RDN and Omni substantially degrade SSIM due to the scarcity of training data.
While optimization-based methods (BIRD and GDP) with the same diffusion prior degrade both PSNR and SSIM, our method achieves clear gains in 3D SR performance, justifying our design choice.

\begin{figure}[h!]
\centering
\includegraphics[width=0.99\textwidth]{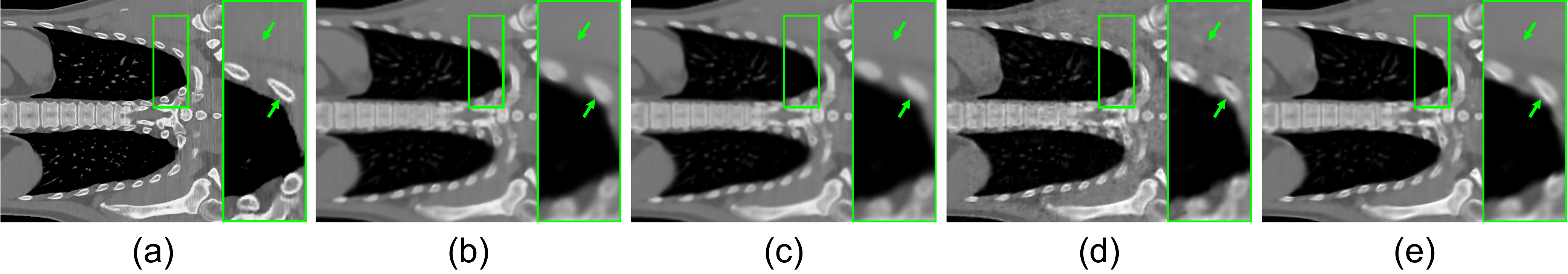}
\centering
\caption{
    % \HJ{thicker box contour width, add yellow contour to side image, revise a,b,c,d,e with the name of methods (a) Ground Truth (b)...}
    Visual comparisons of 3D CT reconstruction across activation functions.
    (a) Ground truth, (b) Softplus, (c) ReLU, (d) Sine, and (e) Ours (PReLU).
    %
    % The yellow box and zoom-in highlight regions where our method excels at reconstruction.
    Ours effectively enhances structural details while suppressing grainy noise.
}
\label{fig:activation_results}
\end{figure}
\noindent\textbf{(2) Effectiveness of NAB-GS.}
\Tabref{tab:ablation_3d} validates NAB-GS.
First, we compare to existing reconstruction methods (FDK~\cite{feldkamp1984practical}, NAF~\cite{zha2022naf}, SAX-NeRF~\cite{cai2024structure}, and R$^2$-GS~\cite{ruyi2024r2_gaussian}).
Specifically, ours outperforms R$^2$-GS in both PSNR ($4\times$: +0.53, $8\times$: +0.71) and SSIM ($4\times$: +0.0118, $8\times$: +0.0116), demonstrating the effectiveness of the residual learning-based strategy.
Second, we compare alternative activations (softplus, ReLU, and sine) with PReLU.
Although the raw density $z$ can take negative values, softplus and ReLU strictly enforce non-negativity, failing to capture boundaries in overestimated regions (\figref{fig:activation_results}(b) and (c)).
Although sine allows non-negative outputs, it introduces grainy artifacts as shown in~\figref{fig:activation_results}(d).
In contrast, PReLU leverages a learnable negative slope to accommodate negative values and achieves the best performance as in~\figref{fig:activation_results}(e).

\section{Conclusion}

% In this paper, we proposed a novel zero-shot 3D CT SR framework that eliminates reliance on HR-LR volumes by reformulating 3D SR as a 2D projection-prior-driven 3D reconstruction task.
% \HJ{
In this paper, we have proposed a novel zero-shot 3D CT SR framework that reformulates volumetric SR as a 3D reconstruction task driven by diffusion-upsampled 2D projection priors.
% }
%
Furthermore, we introduced NAB-GS, which represents the negative densities to learn signed residual fields that enhance structural details.
%
% As a result, our framework achieves SOTA performance, outperforming previous zero-shot methods.
%
% Our framework demonstrated outperforms previous zero-shot methods on both quantitative and qualitative comparisons on two public datasets.
% \HJ{
Our framework outperforms previous zero-shot methods in both quantitative and qualitative results on two public datasets.
% }
%
% Supported by expert clinical feedback, our framework shows potential for real-world application, with future work directed toward further enhancing inter-slice continuity.
% Furthermore, we conducted the expert evaluations on the quality and clinical potential, showing the real-world potential of the proposed method.
% \HJ{
Moreover, an experts’ evaluation highlights the clinical potential of the proposed method, especially at 4$\times$.
% }
% For future work, we will enhance inter-slice continuity for real-world applicability.
% \HJ{
For future work, we will enhance inter-slice continuity for practical clinical use and conduct evaluations on real-world clinical data.
% }
% \HJ{inter-slice consistency between diffusion-generated SR projections}

\begin{credits}
\subsubsection{\ackname}
The authors would like to thank Byungju Chae for his contributions during the early stage of this research. This work was supported in part by the National Research Foundation of Korea under Grant RS-2024-00349697 and Grant RS-2021-NR060143; in part by the Institute for Information and Communications Technology Planning and Evaluation under Grant IITP-2026-RS-2020-II201819; in part by the Technology Development Program funded by the Ministry of SMEs and Startups (MSS), South Korea, under Grant RS-2024-00437796; in part by the National Research Council of Science and Technology (NST) grant funded by Korean Government [Ministry of Science and Information and Communications Technology (MSIT)] under Grant GTL24031-000; and in part by Korea University Grant.

\subsubsection{\discintname}
The authors have no competing interests to declare that are relevant to the content of this article.
\end{credits}

\bibliographystyle{splncs04}
% \bibliographystyle{ieeetr}

\bibliography{template}
\end{document}